\documentclass[nonacm]{acmart}

\usepackage{xspace}
\usepackage{multirow}
\usepackage{amsmath}
\usepackage{caption}
\usepackage{subcaption}
\usepackage{tcolorbox}
\usepackage{hyperref}
\usepackage{color, colortbl}
\definecolor{Gray}{gray}{0.9}

\AtBeginDocument{%
  \providecommand\BibTeX{{%
    \normalfont B\kern-0.5em{\scshape i\kern-0.25em b}\kern-0.8em\TeX}}}

\begin{document}

\makeatletter
\newcommand{\etal}{\emph{et al}.\@\xspace}
\newcommand*{\eg}{\emph{e.g.,}\@\xspace}
\newcommand*{\ie}{\emph{i.e.,}\@\xspace}
\newcommand*{\aka}{\emph{a.k.a.}\@\xspace}
\newcommand*{\fade}{\textsc{Fair}\@\xspace}
\makeatother

\title[Discerning Legitimate Failures From False Alerts]{Discerning Legitimate Failures From False Alerts: A Study of Chromium’s Continuous Integration}

\author{Guillaume Haben}
\affiliation{%
  \institution{SnT, University of Luxembourg}
  \country{Luxembourg}
}
\email{guillaume.haben@uni.lu}

\author{Sarra Habchi}
\affiliation{%
  \institution{SnT, University of Luxembourg}
  \country{Luxembourg}
}
\email{sarra.habchi@uni.lu}

\author{Mike Papadakis}
\affiliation{%
  \institution{SnT, University of Luxembourg}
  \country{Luxembourg}
}
\email{michail.papadakis@uni.lu}

\author{Maxime Cordy}
\affiliation{%
  \institution{SnT, University of Luxembourg}
  \country{Luxembourg}
}
\email{maxime.cordy@uni.lu}

\author{Yves Le Traon}
\affiliation{%
  \institution{SnT, University of Luxembourg}
  \country{Luxembourg}
}
\email{yves.letraon@uni.lu}

\renewcommand{\shortauthors}{Haben et al.}

\begin{abstract}
  Flakiness is a major concern in Software testing. Flaky tests pass and fail for the same version of a program and mislead developers who spend time and resources investigating test failures only to discover that they are false alerts.
  In practice, the \textit{defacto} approach to address this concern is to rerun failing tests hoping that they would pass and manifest as false alerts.
  Nonetheless, completely filtering out false alerts may require a disproportionate number of reruns, and thus incurs important costs both computation and time-wise.
  As an alternative to reruns, we propose \fade, a novel lightweight approach that classifies test failures into false alerts and legitimate failures. 
  \fade relies on a classifier and a set of features from the failures and test artefacts. 
  To build and evaluate our machine learning classifier, we use the continuous integration of the Chromium project.
  In particular, we collect the properties and artefacts of more than 1 million test failures from 2,000 builds. 
  Our results show that \fade can accurately distinguish legitimate failures from false alerts, with an MCC up to 95\%. 
  Moreover, by studying different test categories: GUI, integration and unit tests, we show that \fade classifies failures accurately even when the number of failures is  limited. 
  Finally, we compare the costs of our approach to reruns and show that \fade could save up to 20 minutes of computation time per build.
\end{abstract}




\maketitle

\section{Introduction}

Continuous Integration (CI) is a software engineering process that allows developers to frequently merge their changes in a shared repository~\cite{CI}. 
To ensure a fast and efficient collaboration, the CI automates parts ---if not all--- of the development lifecycle.
Regression testing is an important step of this cycle as it ensures that new changes do not break existing features. 
Test suites are run for every commit and test results report if changes should be integrated to the codebase or not.
Deterministic tests, which have consistent results, are then primordial to keep a smooth automation and a trouble-free CI.
Unfortunately, some tests, commonly named flaky tests, exhibit a non-deterministic behavior as they both pass and fail for the same version of the codebase. 
When flaky tests fail, they send false alerts to developers who are evaluating their code changes.
Indeed, developers spend time and effort investigating these flaky failures, only to discover that they are not due to legitimate faults~\cite{Eck2019}.
When these false alerts occur frequently,  developers may lose trust in their test suites and stop considering failures even if some of them are caused by real faults. 
In this way, false alerts defy the purpose of software testing and hinder the flow of the CI.


Flakiness affects both open source and industrial projects~\cite{Bell2018,Kowalczyk2020,Lam2019iDFlakies}. 
Reports from Google show that almost 16\% of their 4.2 million tests have some level of flakiness. 
This leads them to spend between 2 to 16\% of their computation resources in rerunning flaky tests~\cite{Micco2017}. Many other companies reported having trouble dealing with flakiness, including Microsoft~\cite{Lam2020b}, Spotify~\cite{FlakinessSpotify} and Firefox~\cite{Rahman2018}. 

%

In an attempt to better understand flaky tests, researchers studied the nature and root causes of flakiness~\cite{Lam2019RootCausing,Luo2014,Dutta2020,Thorve2018}. 
These studies showed that the causes differ depending on projects, frameworks, and programming languages.
However, they concurringly report that  asynchronous waits, concurrency, and order-dependency are often the most prevalent root causes~\cite{Gruber2021,romano2021empirical}. 
As for flakiness mitigation, rerunning failing tests is still the main strategy used by practitioners to identify flaky tests~\cite{Eck2019}. If we can observe different outcomes of a test (passing or failing) when rerun on the same version of a program, we know it is a flaky test. But the rerun approach can require a lot of time and compute resources. Studies often re-execute test suites a large number of times, 100 times~\cite{Pinto2020}, 400 times~\cite{Gruber2021} or even 10,000 times ~\cite{alshammari2021flakeflagger} and are still able to uncover unseen flaky tests.
Hence, other tools such as DeFlaker~\cite{Bell2018} and iDFlakies~\cite{Lam2019iDFlakies} were designed to detect flaky tests with a minimal number of reruns. 
Recently, several approaches relied on machine-learning to predict flaky tests based on code vocabulary, code coverage, and dynamic features~\cite{alshammari2021flakeflagger,Pinto2020,Haben2021}, allowing flakiness detection without reruns.
Nevertheless, all these studies focus on distinguishing flaky tests from reliable tests and do not address the distinction between false alerts (\ie flaky failures) and legitimate test failures (\ie real regressions in the code).
%

In this paper, we address the problem of flakiness by focusing on test executions. 
In particular, we propose \fade{} (for FAIlure Recognizer), a novel lightweight approach for discerning legitimate failures from false alerts that are caused by flaky tests. 
Our approach relies on dynamic features (\eg stack-trace and run duration) and static features (\eg test command and code) to discard false alerts while avoiding reruns and coverage computation. 
We then conduct an empirical study on the Chromium project~\cite{TheChromiumProjects} to evaluate the effectiveness and efficiency of our approach.
Our study covers more than 1 million test executions, 2,000 builds, and 41 test suites that include unit, integration, and Graphical User Interface (GUI) tests.
With this study, we aim to answer the following research questions:
\begin{itemize}
    \item \textbf{RQ1:} How effective is \fade{} at identifying legitimate failures?
    \item \textbf{RQ2:} How effective is \fade{}  on different test categories?
    \item \textbf{RQ3:} What features are the most significant for \fade{}?
    \item \textbf{RQ4:} How efficient is \fade{} compared to reruns?
\end{itemize}
Our results show that:
\begin{itemize}
    \item \fade can accurately predict if a failure is legit or a false alert with an MCC up to 95\%.
    \item \fade shows great performances in detecting legitimate failures from GUI tests, with an MCC reaching 98\%. For other test categories (integration and unit), despite a low number of samples to train and test on, \fade is able to decently predict failure classes.
    \item When investigating the most important features used in the model's decisions, the run duration comes first, followed by properties and keywords extracted from the artefacts (stack trace, test source, test command, etc).
    \item Compared to the computation time required by reruns, \fade is faster by an order of 10\textsuperscript{5}.
\end{itemize}

The remaining of this paper is organised as follows. Section~\ref{section:chromium} introduces the chromium CI and our data collection process.
Section~\ref{section:motivation} presents a preliminary analysis of the collected test failures.
Section~\ref{section:approach} explains our failure classification approach \fade{} and Section~\ref{section:evaluation} describes our evaluation protocol.
Section~\ref{section:results} presents the evaluation results and Section~\ref{section:threats} discusses the threats to validity.
Finally, Section~\ref{section:related} discusses related works and Section~\ref{section:conclusion} concludes with the main findings.
For replication purposes, we provide a comprehensive package that includes all the used datasets and scripts~\footnote{\url{https://github.com/GuillaumeHaben/FAIR-ReplicationPackage}}.



\section{The Chromium project}\label{section:chromium}
\subsection{Continuous integration environment}
Started in 2008, with more than 2,000 contributors and 25 million lines of code, the Chromium web browser is one of the biggest open-source projects currently existing. Google is one of the main maintainers, but other companies and contributors also take part in its development.
Chromium relies on LuCI as a CI platform~\cite{onlineChromiumGithub}.
It uses more than 900 parallelized builders, each one of them used to build with different settings (\eg different compilers, instrumented versions for memory error detection, fuzzing, etc)  and to target different operating systems (\eg Android, MacOS, Linux, and Windows). 
Each builder is responsible for a list of builds triggered by each commit made to the project. In a build, we find details about build properties, start and end times, status (\ie success or failure), a listing of the steps and links to the logs. 

In the beginning, building and testing were sequential. Builders used to compile the project and zip the results to builders responsible for tests. Testing was taking a lot of time, slowing developers productivity and testing Chromium for several platforms was not conceivable, even by adding more computation power. A swarming infrastructure was then introduced in order to scale according to the Chromium development team productivity, to keep getting the test results as fast as possible and independently from the number of tests to run or the number of platforms to test. A fleet of 14,000  build bots is currently available to run tasks in parallel. This setup helps running tests with low latency and sustains hundreds of commits per day~\cite{TheChromiumProjects}.

In this study, we focus on testers, \ie builders only responsible for running tests. At the time of writing, we found 47 testers running Chromium test suites on different versions of operating systems. About 300,000 tests are divided into 41 test suites, the biggest ones being \textit{blink\_web\_tests} (testing the rendering engine) and \textit{base\_unittests} with more than 60,000 tests each.
For each build, we can obtain information about test results. The test is labelled as \textsf{pass} when it successfully passed after one execution. In case of a failure, \textit{LuCI} automatically reruns the test from 2 to 5 times depending on configurations. If all reruns fail, the test is labelled as \textsf{unexpected} and will trigger a build failure. If a test passes after having one or more failed executions during the same build, it is labelled as \textsf{flaky} and won't prevent the build from passing. 
In the remaining, we refer to \textsf{unexpected} failures as \textsf{legitimate} failures since their failing behaviour persisted through several reruns.
On the other hand, we refer to \textsf{flaky} failures as \textsf{false alerts}.



\subsection{Data Collection}
We focus our data collection on the two testers \textit{Linux Tests} and \textit{Win10 Tests x64}, which are responsible for executing tests of the two main operating systems supported by Chromium. 
We collected the records of 1,000 builds for each tester. This represents 45 days of development taken between February and March 2021. 
By querying \textit{LuCI}'s API, we retrieved the details of each test failure that occurred within these builds. 
Table ~\ref{table:infoDataset} reports the number of false alerts and legitimate failures found in this dataset. 
We list the total number of runs that fall into each category and the number of unique failing tests. 
Interestingly, we observe that the number of failures that turn out to be false alerts is much larger than the legitimate ones, 969,417 and 225,762 respectively.
That is, false alerts represent 81\% of the failures in the Chromium CI, whereas legitimate failures only represent 19\%.
We also observe that the number of tests responsible for failures is very low compared to the number of failures.
This indicates that the same tests keep failing across different builds. 

\begin{table*}[t]
\vspace{1.0em}
\begin{center}
\caption{Data collected from the Chromium CI}
\label{table:infoDataset}
\begin{tabular}{l | c | c c | c c } 
\hline
\multirow{2}{*}{\textbf{Tester}} & \multirow{2}{*}{\textbf{\#Builds}} & \multicolumn{2}{c|}{\textbf{False alerts}} & \multicolumn{2}{c}{\textbf{Legitimate failures}} \\
{} & {} & \textbf{\#Tests} & \textbf{\#Failures} &  \textbf{\#Tests} & \textbf{\#Failures} \\
\hline 
Linux Tests & 1,000 & 7,063 & 191,886  & 2,855 & 175,904 \\
Win10 Tests x64 & 1,000 & 40,668 & 777,531 & 1,482 & 49,858 \\
\hline
Total & 2,000 & 47,731 & 969,417 & 4,337 & 225,762 \\
\hline
\end{tabular}
\end{center}
\vspace{1.0em}
\end{table*}

\begin{tcolorbox}
False alerts represent more than 81\% of the failures detected in the Chromium CI, whereas legitimate failures only represent 19\%.
\end{tcolorbox}

\section{Preliminary Analysis}\label{section:motivation}
This section presents our preliminary analysis of test failures in the Chromium CI.
The objective is to study the characteristics of test failures, both false alerts and legitimate failures, and pave the way for our approach design.
In accordance with the study objective, we formulate the following preliminary question:
\begin{itemize}
    \item \textbf{PQ:} How do legitimate failures compare to false alerts?
\end{itemize}
To answer this question, we compare false alerts and legitimate failures in terms of:
\begin{itemize}
    \item \textbf{History:} This represents the historical behaviour (pass, failure, flake) of tests responsible for failures. 
    \item \textbf{Duration:} This represents the duration of test executions leading to legitimate failures and false alerts.
\end{itemize}

\subsection{Test history}
We analyse the history of tests responsible for legitimate failures and false alerts by relying on the metrics \textsf{flakeRate} and \textsf{failRate}.
For a failure of a test $t$ occurring at a commit $c_{n}$, we analyse all the commits from a time window $w$ (\ie from $c_{n-w}$ to $c_{n-1})$ to calculate the rates as follows: \\

\noindent\begin{minipage}{.5\linewidth}
\begin{equation}
  flakeRate(t,n) = \frac{ \sum_{x=n-w}^{n-1} flake(t,x) } {w}
\end{equation}
\end{minipage}%
\begin{minipage}{.5\linewidth}
\begin{equation}
  failRate(t,n) = \frac{ \sum_{x=n-w}^{n-1} fail(t,x) } {w}
\end{equation}
\end{minipage}
\\

where $flake(t,x) = 1$ if the test $t$ flakes in the commit $c_{x}$ and null otherwise, while $fail(t,x) = 1$ if the test $t$ fails persistently (for five reruns) in the commit $c_{x}$ and null otherwise.
These two metrics allow us to understand if the history of a test can help in distinguishing false alerts from legitimate failures.
The test execution history (\aka heartbeat) has been used in multiple studies (especially industrial ones~\cite{Kowalczyk2020,LeongSPTM19}) to detect flaky tests.
These studies assume that many flaky tests have distinguishable failure patterns over commits and hence can be detected by observing their history.
In this question, we check whether this assumption holds in the case of Chromium.

\begin{figure}[!htbp]
  \centering
  \begin{minipage}[b]{1\textwidth}
    \includegraphics[width=\textwidth]{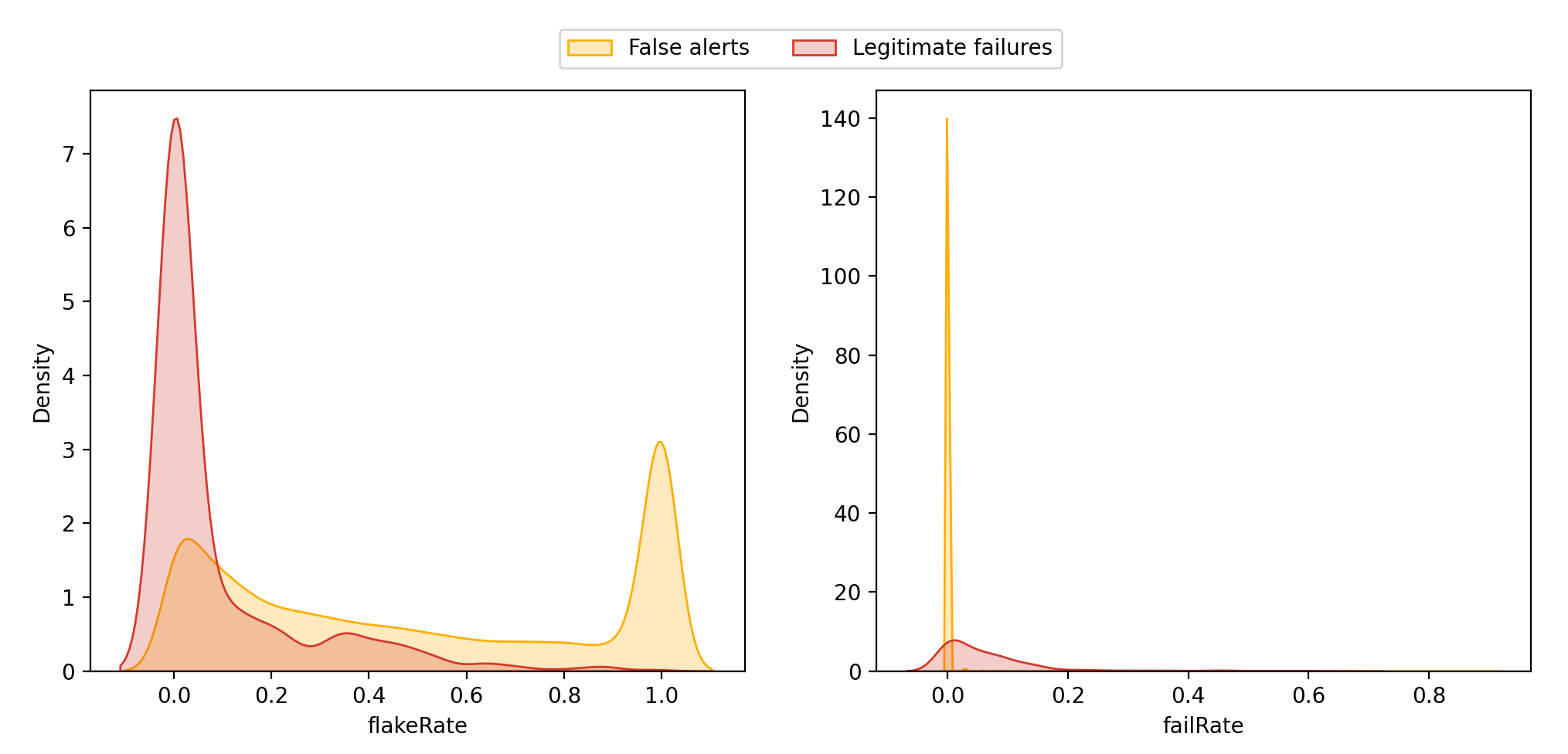}
    \caption{Flake rate (left) and failure rate (right) for runs marked as false alerts and legitimate failures by reruns}
    \label{fig:kdeRates}
  \end{minipage}
\end{figure}

To illustrate differences in rates between false and legitimate failures, we plot the \textsf{flakeRate()} and \textsf{failRate()} for each failure category in Figure~\ref{fig:kdeRates}. 
The flakiness and failure rates are computed using a window of 35 commits.
For false alerts, analyzing their history of failures shows that in most cases, the tests did not have a history of failures. 
The percentage of false alerts having a $failRate() > 0$ is in fact $0.2\%$. 
In contrast, failures marked as false alerts were often witnessed in tests that had false alerts in previous commits. Indeed, only 16\% had a $flakeRate() = 0$ in the past commits. 
In the case of legitimate failures, we observe that they generally occur in tests that did not fail or flake before. 
60\% of legitimate failures never flaked in previous commits. 
From these observations, we may suggest that the \textsf{flakeRate()} can be used to further distinguish between false alerts and legitimate failures.
Nevertheless, there is still an important overlap between the history of false alerts and legitimate failures. 
In particular, in the history of Linux and Windows testers, we observe that:
\begin{itemize}
    \item 16\% of false alerts have $flakeRate() = 0\ and\ failRate() = 0$;
    \item And 19\% of legitimate failures have $flakeRate() = 0\ and\ failRate() = 0$.
\end{itemize}
These two groups constitute a set of failures that have a clean history and thus are impossible to classify into false alerts or legitimate failures based on flakiness and failure metrics.
Hence, we conclude that the historical differences between false alerts and legitimate failures are not sufficient for distinguishing them.

Another interesting observation from figure ~\ref{fig:kdeRates} is that there is a set of legitimate failures that occur in tests that manifested false alerts ($flakeRate() > 0$) in recent commits.   
We suspect that these outliers are false alerts that were not properly classified by reruns (we call them \textit{unreliable failures}).
Indeed, while reruns can affirm that a failure is due to flakiness by manifesting a test pass and fail for the same version, they do not allow us to affirm that a failure is legitimate.
A previous study showed that up to 10,000 reruns can be required to discover flaky tests that have a low flake rate~\cite{alshammari2021flakeflagger}. 
Hence, a legitimate failure in the case of Chromium can still be a false alert (flaky failure) that was not rerun enough to manifest.
This explains the behaviour of these tests which fail and flake intermittently in close time windows.

\subsection{Run duration}
The duration of test failures is one of the features easily retrievable in the Chromium CI. 
Thus, as part of our preliminary analysis, we compare false alerts and legitimate failures in terms of run duration.
For this, we analyse the run duration of all failures from the \textit{Linux Tests} tester. 
Table~\ref{table:duration} shows the distribution of run duration for legitimate and false failures. 
We observe that legitimate failures tend to take longer than false alerts. 
In particular, the median duration of test executions is 1.09 seconds for legitimate failures, but only 0.41 seconds for false alerts.
In the same vein, 25\% of legitimate failures lasted more than 2.11 seconds, whereas the same portion of false alerts only lasted more than 1.31 seconds.
This suggests that some failure properties such as run duration can help us distinguish legitimate failures from false alerts.

\begin{table*}[t]
\vspace{1.0em}
\begin{center}
\caption{Execution duration for test failures in seconds}
\label{table:duration}
\begin{tabular}{ l c c c} 
\cline{2-4}
\multicolumn{1}{c}{} & \textbf{Q1} & \textbf{Median} & \textbf{Q3} \\ \hline
\textbf{False alerts} & 0.18 & 0.41 & 1.31 \\\hline
\textbf{Legitimate failures} &  0.77 & 1.09 & 2.11 \\

\bottomrule
\end{tabular}
\end{center}
\vspace{1.0em}
\end{table*}

\begin{tcolorbox}
False alerts occur recurrently in the same tests, whereas legitimate failures occur in tests that have low chances to fail or flake in the past. 
Yet, due to a large number of failures with a clean history, these differences are not sufficient to discriminate legitimate failures from false alerts.
With regard to run properties, false alerts manifest a shorter run duration than legitimate failures.
\end{tcolorbox}
\section{\fade{}}\label{section:approach}

Our main objective is to distinguish legitimate failures from false alerts efficiently and effectively.
To achieve this, we propose a learning model that relies on dynamic and static properties to classify failures without performing any reruns. 
In the following, we present our feature extraction process and model design.

\subsection{Features}
\paragraph{Data cleaning}
Starting from the dataset retrieved from the Chromium CI, we applied two filters to ensure the quality of our dataset.
First, we excluded some failures that occur in a period where a big chunk of the test suite was failing oddly. A commit undoing changes that introduce these test suite failures was performed after investigation.
Specifically, in \textit{Linux Tests}, for 15 consecutive builds (98177 to 98192), 2,700 tests were failing repeatedly and this results in a total of more than 170,000 failures. 
During the same period, in \textit{Win10 Tests x64}, 1,300 tests were failing per build, resulting in more than 45,000 failures spread across 9 builds (53760 to 53768). 
This irregular amount of failures only impacted JavaScript tests and the root cause remains unexplained.
Therefore, we decided not to consider these builds in our experiments, to avoid a bias towards JavaScript tests.
As a second cleaning step, we removed all the failures that were deemed unreliable based on their history (cf. Section~\ref{section:motivation}).
This filter excluded $0.13\%$ of the failures in our dataset.
\paragraph{Feature extraction}
To construct our features, we analyse the failure properties, \eg run duration, and all the available artefacts.
Depending on the test, different artefacts are potentially available for each failure. Crash logs, stack traces, and the standard error provide various information about the raised exceptions and the execution of the program under test. 
For some tests, we can also retrieve the command used to launch the test.
To widen the feature space, we also retrieve the source code of the test in its current commit using Google Git\footnote{\url{https://chromium.googlesource.com/}}. 
In the case of the test source code, we only retrieved it when the test file was only containing one test. Chromium's codebase --- and tests --- are written in different languages, but the vast majority are HTML and JavaScript tests. As other tests are not easily parsable, we decided to only add the test source to the runs of HTML and JavaScript tests. 
As all these artefacts are of textual nature (crashLog, stackTrace, stderr, and testSource), we build a bag of words~\cite{goldberg2017neural} representation for each of them and we consider them as separate features. When counting words, we use Term Frequency-Inverse Document Frequency (TF-IDF), which aims at lowering the weight on common words present in all documents.
In order to denote the presence of artefacts (or not) and their size, we add the length of all artefacts to the features.
Additionally, we extract the run duration, run status, and run tag status to complete our list of features. Run duration is the execution time needed to run the test. Run status gives information about the run result (\eg passing, failing, and skipped) and run tag status returns more precise information about the result of a run depending on the type of test or test suite (\eg timeout and failure on exit).
Table~\ref{table:infoFeatures} provides a summary of all the extracted features.


\begin{table*}[t]
\vspace{1.0em}
\begin{center}
\caption{Description of the features}
\label{table:infoFeatures}
\begin{tabular}{ l | l } 
\hline
\textbf{Feature name} & \textbf{Feature description} \\
\hline 
command & The vocabulary of the command used to run the test \\
commandLength & The number of characters present in the command artieact \\
crashlog & The vocabulary of the crash log resulting from an error \\
crashlogLength & The number of characters present in the crashlog artefact \\
runDuration & The time spent for this run execution \\
runStatus & 0: ABORT, 1: FAIL, 2: PASS, 3: CRASH, 4: SKIP \\
\multirow[t]{2}{*}{runTagStatus} & 0: CRASH, 1: PASS, 2: FAIL, 3: TIMEOUT, 4: SUCCESS, \\
& 5: FAILURE, 6: FAILURE\_ON\_EXIT, 7: NOTRUN, 8: SKIP, 9: UNKNOWN \\
stackTrace & The vocabulary of the stack trace resulting from an error \\
stackTraceLength & The number of characters present in the stackTrace artefact \\
stderr & The vocabulary of the standard error captured after the test execution \\
stderrLength & The number of characters present in the stderr artefact \\
testSource & The vocabulary of the source code for the test \\
testSourceLength & The number of characters present in the testSource artefact \\
\hline
\end{tabular}
\end{center}
\vspace{1.0em}
\end{table*}

\subsection{Failure classifier}
Following previous studies on flakiness prediction~\cite{Haben2021, Pinto2020} finding that Random Forest yields the best performances in flakiness classification tasks, we rely on this model for our classification as well. Selecting the model that yields the best performance is not in the scope of our study. We rather focus on understanding if the approach is feasible. 
With the two sets of data collected from Linux and Windows testers, we proceed as follows. 
Each failure in our dataset is denoted as an n-dimensional feature vector $X = (x_1,...,x_n)$ where $x_i$ represents one feature. $y = \{0,1\}$ indicates if the failure is from the false alert class (0) or from the legitimate failure class (1).
Once all vectors are created, we randomly split our dataset by including 80\% of it in the training set and 20\% in the test set, conserving the class ratio in each subset (stratified).
Using the training set, we search for the optimal set of hyperparameters for our Random Forest Classifier. To do so, we apply a 5-fold cross-validation technique using randomized search \cite{bergstra2012random}. 
Considering our dataset, this step takes between 5 to 30 minutes depending on the test categories considered (See RQ2) on a machine with Intel Core 2.3GHz 8‑core 9th‑generation and 32-GB RAM.
Once the optimal hyperparameters are tuned, we then search for the best threshold for precision and recall. We avoid using ROC curves which tend to be overly optimistic in their view of algorithms' performance in presence of a high level of imbalanced data. Instead, we use the precision-recall curve recommended as an alternative in this case~\cite{precisionRecallCurve}. Tuning the behaviour of a model by calibrating its classification threshold enables us to find the best trade-off between precision and recall.
Finally, we refit the model on the whole training set and we evaluate its performance on the test set. 
The classified failures are categorized as:
\begin{itemize}
    \item True Positives (TP): failures that are correctly classified as legitimate failures;
    \item False Positives (FP): false alerts that are wrongfully classified as legitimate;
    \item False Negatives (FN): legitimate failures that are wrongfully classified as false alerts;
    \item True Negatives (TN): failures correctly classified as false alerts.
\end{itemize}



\section{Evaluation}\label{section:evaluation}
This section presents our experimental protocol for evaluating \fade{} and answering our research questions.

\subsection{RQ1: How effective is \fade{} at identifying legitimate failures?}
In this question, we evaluate the performances of \fade{} at detecting legitimate failures.
The objective is to investigate the possibility of using a machine learning model and failure properties to perform this detection. 
To evaluate our model, we rely on the following metrics:
\begin{itemize}
    \item Precision: Given by:  $\frac{TP}{TP+FP}$.
    \item Recall: Given by: $ \frac{TP}{TP+FN}$.
    \item F1-score: Given by:  $2 * \: \frac{precision\: * \: recall}{precision+recall}$
    \item MCC: The accuracy of a model is sensitive to the class imbalance. In particular, the precision and recall metrics can easily be impacted when one class is underrepresented, which is the case for our dataset. To alleviate this issue, we report the Matthews Correlation Coefficient (MCC) which is a more reliable statistical rate to avoid over-optimistic results in the case of an imbalanced dataset \cite{chicco2020advantages}.
    This metric takes into consideration all four entries of the confusion matrix.
    MCC is given by:

    \[
    \frac{TN \times TP - FP \times FN}{\sqrt{(TN+FN)(TP+FP)(TN+FP)(FN+TP)}}
    \]
\end{itemize}


\subsection{RQ2: How effective is \fade{} on different test categories?}
In this question, we evaluate the performances of \fade{} on different test categories.
In heavily tested projects, different types of tests are used to assess the system quality.
In the case of Chromium, test suites can be classified into three test categories: Graphical User Interface (GUI), integration, and unit tests.
Based on their purpose, these tests differ in terms of scope and cost.
In particular, unit tests target lower granularity components (\eg classes or functions) and their execution is generally cheap both computation and time-wise.
On the other hand, GUI tests evaluate the system as a whole (end-to-end) and consume more time and resources.
Based on these differences, we expect these categories of tests to manifest different execution characteristics, which may influence our false alert detection approach.
Hence, we take into account the test categories and evaluate the performances of \fade{} with two strategies:
\begin{itemize}
    \item \textbf{Cross-category evaluation:} In this strategy, we train the model using data from the different categories of tests, following their distribution in the original dataset. Then, we evaluate the model performances on three separate validation sets: GUI tests, integration tests and unit tests.
    This strategy allows a fine-grained evaluation of our approach and shows its performances on the three test categories.
    
    \item \textbf{Intra-category evaluation:} We separately build and evaluate one model per test category. 
    This allows us to assess the performances of our strategy when only one test category is available.
    
\end{itemize}

Table~\ref{table:infoTestSuites} provides details about Chromium's test suites. 
For each test suite, we report the number of collected \textsf{false alerts}, \textsf{legitimate failures}, and the tests that are concerned by these failures. 
For the sake of readability, we only present test suites for which at least 100 failures are collected. 
The two largest test suites are \textit{blink\_web\_tests} and \textit{webkit\_unit\_tests}. Consequently, most failures come from these test suites.

\begin{table*}[t]
\vspace{1.0em}
\begin{center}
\caption{Details about Chromium test suites for the considered builds}
\label{table:infoTestSuites}
\resizebox{\textwidth}{!}{
\begin{tabular}{l | l | c c c | c c c } 
\hline
\multirow{1}{*}{\textbf{Test suite}} & \multirow{1}{*}{\textbf{Test type}} & \multicolumn{3}{c|}{\textbf{Win10 Tests x64}} & \multicolumn{3}{c}{\textbf{Linux Tests}} \\
{} & {} & \#Tests & \#False alerts & \#Legitimate failures & \#Tests & \#False alerts & \#Legitimate failures  \\
\hline 
accessibility\_unittests & unit & 0 & 0 & 0 & 1 & 961 & 0 \\
aura\_unittests & unit & 0 & 0 & 0 & 74 & 318 & 0 \\
blink\_platform\_unittests & unit & 3 & 81 & 0 & 3 & 204 & 0 \\
blink\_web\_tests & GUI & 26,137 & 190,489 & 504 & 1,389 & 74,566 & 584 \\
browser\_tests & integration & 1,368 & 11,228 & 208 & 2,529 & 20,297 & 222 \\
components\_unittests & unit & 27 & 990 & 12 & 100 & 601 & 8 \\
content\_browsertests & integration & 241 & 6,380 & 40 & 268 & 583 & 28 \\
content\_unittests & unit & 80 & 209 & 0 & 3 & 6 & 0 \\
cronet\_tests & integration & 2 & 206 & 0 & 0 & 0 & 0 \\
elevation\_service\_unittests & unit & 1 & 968 & 0 & 0 & 0 & 0 \\
events\_unittests & unit & 2 & 164 & 0 & 0 & 0 & 0 \\
extensions\_unittests & unit & 15 & 5,018 & 0 & 0 & 0 & 0 \\
gcp\_unittests & unit & 46 & 3,889 & 0 & 0 & 0 & 0 \\
interactive\_ui\_tests & GUI & 65 & 182 & 18 & 144 & 486 & 74 \\
jingle\_unittests & unit & 1 & 327 & 0 & 0 & 0 & 0 \\
media\_blink\_unittests & unit & 3 & 11 & 0 & 3 & 428 & 2 \\
mojo\_unittests & unit & 11 & 131 & 0 & 0 & 0 & 0 \\
net\_unittests & unit & 27 & 452 & 2 & 11 & 881 & 0 \\
non\_skia\_renderer\_swiftshader\_blink\_web\_tests & GUI & 0 & 0 & 0 & 12 & 210 & 0 \\
not\_site\_per\_process\_blink\_web\_tests & GUI & 0 & 0 & 0 & 1,486 & 60,684 & 680 \\
perfetto\_unittests & unit & 0 & 0 & 0 & 19 & 410 & 0 \\
pixel\_browser\_tests & GUI & 100 & 633 & 72 & 0 & 0 & 0 \\
remoting\_unittests & unit & 12 & 259 & 0 & 0 & 0 & 0 \\
services\_unittests & unit & 21 & 698 & 0 & 10 & 386 & 0 \\
setup\_unittests & unit & 1 & 135 & 0 & 0 & 0 & 0 \\
shell\_dialogs\_unittests & unit & 4 & 1,922 & 0 & 0 & 0 & 0 \\
sync\_integration\_tests & integration & 93 & 133 & 0 & 91 & 103 & 2 \\
telemetry\_unittests & unit & 0 & 0 & 0 & 2 & 7,283 & 0 \\
unit\_tests & unit & 161 & 2,703 & 14 & 154 & 2,285 & 26 \\
views\_unittests & unit & 0 & 0 & 0 & 307 & 673 & 0 \\
vulkan\_swiftshader\_blink\_web\_tests & GUI & 0 & 0 & 0 & 14 & 2,898 & 0 \\
webkit\_unit\_tests & unit & 11,959 & 528,703 & 4 & 75 & 10,350 & 4 \\
wm\_unittests & unit & 0 & 0 & 0 & 59 & 298 & 0 \\
\hline
\end{tabular}
}
\end{center}
\vspace{1.0em}
\end{table*}

\subsection{RQ3: What features are the most significant for \fade{}?}
In this question, we shed light on the main features used by our model to detect legitimate failures and false alerts.
To identify these features, we compute the information gain provided by each feature and we sort them accordingly. The information gain is based on the Gini coefficient of the features.
Afterwards, we report and analyse the ten features that provide the most informational gain.

\subsection{RQ4: How efficient is \fade{} compared to reruns?}
In the Chromium project, the current approach to deal with false alerts is to rerun tests to observe flakiness.
In this question, we compare our approach \fade{} with reruns as a baseline, in terms of execution time required to classify a failure.  
To perform this comparison, we extract from LuCI:
\begin{itemize}
    \item The average time required by reruns to identify a failure as a false alert;
    \item The average time required by reruns to identify a legitimate failure;
\end{itemize}
Based on these values, we compute the total time required to classify all failures in our dataset and estimate the rerun cost per build.
For the costs of \fade{}, we report the cost of model training and the average time consumed to classify a failure.
Thence, we deduce the overall cost for classifying all the failures of our dataset and the average cost of \fade{} per build.
\section{Results}\label{section:results}

\subsection{RQ1: How effective is \fade{} at identifying legitimate failures?}

Table ~\ref{table:modelResultsOnTestTypes} presents the precision, recall, F1-score, and MCC score of \fade{} when evaluated with different strategies.
In this question, we focus on the results for All $\rightarrow$ All, which show the model performances when trained and tested on the whole dataset, regardless of test categories.
For the Linux tester, we observe that the model has great performances with an MCC of 95\%.
\fade{} detects legitimate failures with a precision of 98\% and a recall of 93\%.
Similar results are observed in the Windows tester, with a precision of 95\% and a recall of 92\%.
Furthermore, the MCC and F1-score are also stable at 94\%.
Hence, we can conclude that \fade{} distinguishes accurately legitimate failures from false alerts.

\begin{tcolorbox}
\fade{} shows great performances in discriminating legitimate failures from false alerts, with a minimum MCC and F1-score of 94\%.
\end{tcolorbox}

\subsection{RQ2: How effective is \fade{} on different test categories?}

The remaining of Table~\ref{table:modelResultsOnTestTypes} reports the performance of \fade{} on specific test categories.
First, we analyse the results of the cross-category evaluation.
This represents the cases where the model is trained using data from different categories but tested on a specific test category, \ie All $\rightarrow$ GUI, All $\rightarrow$ Integration, and All $\rightarrow$ Unit.
The results show that training on a diverse dataset still allows the model to accurately detect legitimate failures from different test categories.
GUI and integration tests have the best performances with an MCC score ranging between 87\% and 98\% in the Linux dataset, and between 91\% and 94\% in the Windows dataset.
Unit tests show slightly lower results with an MCC of 75\% for both Linux and Windows datasets.
These positive results show that the observations of RQ1 were not solely based on the performances on GUI tests (the predominant category in our datasets), and the model can distinguish legitimate failures for the three test categories.

Secondly, we analyse the results of intra-category evaluation, \ie GUI $\rightarrow$ GUI, Integration $\rightarrow$ Integration, and Unit $\rightarrow$ Unit.
For GUI tests, we observe great performances in the two datasets.
The minimum precision and recall are 96\% and 93\% respectively, and the MCC and F1-score reach 98\% in the Linux dataset.
These performances decrease slightly for the other two test categories, integration and unit, especially in the Linux dataset.
Specifically, for integration tests, \fade{} has an MCC and F1-score between 78\% and 94\%, and precision and recall between 76\% and 96\%.
For unit tests, the performances have more variations as MCC and F1-score range between 70\% and 84\%.
The decrease and variation for integration and unit tests are due to the limited number of legitimate failures from this category.
In particular, the number of legitimate unit test failures is 18 and 35 in the Linux and Windows datasets respectively.
This means that the training sets contain only 14 and 28 legitimate failures (80\% of the total number). 
These limited samples do not allow the model to observe enough examples of the target class (legitimate failures) and hinder its capabilities. 
However, the overall decent results for all categories and the great results for GUI tests suggest that our approach is capable of detecting legitimate failures from different test categories.

\begin{tcolorbox}
\fade{} can accurately detect legitimate failures that arise from GUI, integration, and unit tests. When provided with large datasets, \fade{} can reach great performances with an MCC of 98\%, and even with very limited learning samples (only 14 legitimate failures), its performances remain decent with a minimum MCC of 70\%.
\end{tcolorbox}


\begin{table*}[t]
\vspace{1.0em}
\begin{center}
\caption{\fade{} evaluation with different strategies}
\label{table:modelResultsOnTestTypes}
\resizebox{\textwidth}{!}{
\begin{tabular}{l | l | c c c c | r r r} 
\hline
\multirow{2}{*}{\textbf{Dataset}} & \multirow{2}{*}{\textbf{Training strategies}} & \multicolumn{4}{c|}{\textbf{Metrics}} & \multicolumn{3}{c}{\textbf{Classes}} \\
{} & {} & Precision & Recall & F1 & MCC & False alerts & Legit failures & Total \\
\hline 
\multirow[c]{3}{*}{Linux Tests} & All $\rightarrow$ All & 98,0\% & 93,7\% & 95,8\% & 95,8\% & 182,989 & 1,274 & 184,263 \\
\cline{2-9}

& All $\rightarrow$ GUI & 99,6\% & 97,4\% & 98,4\% & 98,4\% & 182,989 & 1,274 & 184,263 \\
& All $\rightarrow$ Integration & 78,6\% & 97,1\% & 86,8\% & 87,2\% & 182,989 & 1,274 & 184,263 \\
& All $\rightarrow$ Unit & 75,0\% & 75,0\% & 75,0\% & 75,0\% & 182,989 & 1,274 & 184,263 \\
\cline{2-9}

& GUI $\rightarrow$ GUI & 99,5\% & 97,8\% & 98,7\% & 98,6\% & 137,749 & 1,124 & 138,873 \\
& Integration $\rightarrow$ Integration & 80,0\% & 76,9\% & 78,4\% & 78,3\% & 20,746 & 132 & 20,878 \\
& Unit $\rightarrow$ Unit & 100\% & 50,0\% & 66,7\% & 70,7\% & 24,490 & 18 & 24,508
\\
\hline 
\multirow[c]{3}{*}{Win10 x64} & All $\rightarrow$ All & 95,8\% & 92,9\% & 94,4\% & 94,4\% & 744,545 & 495 & 745,040 \\
\cline{2-9}

& All $\rightarrow$ GUI & 93,7\% & 95,2\% & 94,4\% & 94,4\% & 744,545 & 495 & 745,040 \\
& All $\rightarrow$ Integration & 93,1\% & 90,0\% & 91,5\% & 91,5\% & 744,545 & 495 & 745,040 \\
& All $\rightarrow$ Unit & 100\% & 57,1\% & 72,7\% & 75,6\% & 744,545 & 495 & 745,040 \\
\cline{2-9}

& GUI $\rightarrow$ GUI& 96,7\% & 93,5\% & 95,1\% & 95,1\% & 189,347 & 310 & 189,657 \\
& Integration $\rightarrow$ Integration & 96,6\% & 93,3\% & 94,9\% & 94,9\% & 17,722 & 150 & 17,872 \\
& Unit $\rightarrow$ Unit & 100\% & 71,4\% & 83,3\% & 84,5\% & 537,476 & 35 & 537,511 \\
\hline
\end{tabular}
}
\end{center}
\vspace{1.0em}
\end{table*}

\subsection{RQ3: What features are the most significant for \fade{}?}
Table~\ref{table:RQ3} reports the most significant features for our model.
Column \textit{Gain} presents the informational gain provided by each feature and column \textit{origin} presents their sources.
Overall, we observe that the model relies on features from both run properties and artefacts.
The run duration is the most important feature with an information gain of 0.06.
This confirms the results of our preliminary analysis where legitimate failures showed a tendency of longer run durations than false alerts.
Another run property that influences the model is the runTagStatus, with an informational gain of $0.01$.
This property hints at the failure cause like crash, timeout, and failure on exit.
From our analysis, we observed that false alerts tend to be more related to timeouts than legitimate failures.
Indeed, 15\% of false alerts are caused by timeouts, whereas only 4\% of legitimate failures are related to timeouts.
For features extracted from artefacts, the length of the stack trace, the test source, and the execution command show an important influence on the model.
This aligns with previous industrial observations, which indicate that the test length is positively correlated to flakiness~\cite{alshammari2021flakeflagger}.
Besides the length, the vocabulary of artefacts plays an important role in the classification, with five features in the list.

\begin{tcolorbox}
\fade{} capitalizes on dynamic properties like run duration and failure status to discern legitimate failures. Properties and keywords from artefacts, like the stack trace and test command, also play a significant role in the classification.
\end{tcolorbox}

\begin{table*}[t]
\begin{center}
\caption{Most significant features}
\label{table:RQ3}
\resizebox{0.5\textwidth}{!}{
\begin{tabular}{c c c c c c} 
\toprule
\textbf{Feature} & \textbf{Gain} & \textbf{Origin} \\ 
\hline
runDuration & 0.061 & Run properties \\
stackTraceLength & 0.027 & Artefacts \\
elements & 0.019 & Artefacts vocabulary \\
testSourceLength & 0.014 & Artefacts \\
update & 0.012 & Artefacts vocabulary \\
next & 0.011 & Artefacts vocabulary \\
commandLength & 0.010 & Artefacts \\
load & 0.010 & Artefacts vocabulary \\
runTagStatus & 0.010 & Run properties \\
portal & 0.008 & Artefacts vocabulary \\
\bottomrule
\end{tabular}
}
\end{center}
\end{table*}

\subsection{RQ4: How efficient is \fade{} compared to reruns?}

Table \ref{table:cost-reruns} reports the time required by reruns to classify failures. 
On average, reruns take 2.3 seconds to observe flakiness and confirm that a failure is a false alert, while it takes less than one second to identify a legitimate failure.
Reruns require 686 hours to classify all the failures in our dataset of 1,195,179 failures.
Considering that these failures come from 2,000 builds, this represents a cost of around 20 minutes per build.

Table~\ref{table:cost-fade} reports the time required by \fade{} to classify failures.  
Regardless of its type, \fade{} requires 0.001 milliseconds to classify a failure.
Hence, it takes 2.2 seconds to classify all the 1,195,179 failures and consumes on average 1.1 milliseconds per build.
In comparison, \fade{} has a clear edge in terms of efficiency.
Using \fade{}, the individual cost of failure classification goes down from one second to 0.001 milliseconds (900,000 times faster), and the cost per build goes from 20 minutes to 1.1ms.

\begin{table*}[t]
\begin{center}
\caption{The cost of reruns used to detect legitimate failures}
\label{table:cost-reruns}
\begin{tabular}{c c c c c c} 
\toprule
\multicolumn{2}{c}{\textbf{False Alerts}} & \multicolumn{2}{c}{\textbf{Legit Failures}} &
\multicolumn{2}{c}{\textbf{Rerun cost}} \\
Average & \#Instances & Average & \#Instances & Per build& Total \\\hline
0.41 seconds & 969,417 & 1.09 seconds & 225,762 & 20 minutes & 686 hours \\
\bottomrule
\end{tabular}
\end{center}
\end{table*}

\begin{table*}[t]
\begin{center}
\caption{The cost of building and using \fade{}}
\label{table:cost-fade}
\begin{tabular}{c c c c c} 
\toprule
\multicolumn{1}{c}{\textbf{Training}} & \multicolumn{2}{c}{\textbf{Prediction}} &
\multicolumn{2}{c}{\textbf{Overall cost}} \\
Cost & Average & \#Instances & Per build & Total\\\hline
<30 minutes & 0.001 milliseconds & 1,195,179  & 1.1 milliseconds & 2.2 seconds \\
\bottomrule
\end{tabular}
\end{center}
\end{table*}

\begin{tcolorbox}
\fade{} is able to classify failures 900,000 times faster than reruns. Using \fade{}, the cost of failure classification per build will go down from 20 minutes to one millisecond.
\end{tcolorbox}
\section{Threats to validity}\label{section:threats}

\subsection{External validity} 
One possible threat to our external validity is the generalizability to other software development infrastructures.
We rely solely on data from the Chromium project, which has a specific organization with a continuous integration setup that may not generalize to other software projects. 
Thus, some of the features used in this study (\eg run tag status) may be difficult to obtain, impacting the model performances.
We encourage future studies to investigate other dynamic features and testing artefacts to further investigate their usability in discerning false alerts.
Another threat to the generalizability is the variety of test categories.
To the best of our knowledge, this is the first study on false alert detection that considers multiple test categories.
Nonetheless, as our categories are limited to GUI, integration, and unit tests, investigations are yet to be done to check if results hold for other categories of tests. 
Finally, the choice of  programming languages and testing environments may also affect the validity of our results.
We perform our experiments on two builders responsible for running test suites on two different operating systems and the analyzed tests are written in different programming languages.
Yet, to get better insights on the approach's efficiency and effectiveness, we encourage investigations on different configurations.

\subsection{Construct validity} 
A potential threat to our study's validity is the definitions of legitimate failures.
We build our dataset from the Chromium CI, which relies on reruns to classify failures.
Consequently, the validity of our results may be hindered by misclassifications present in the CI.
To alleviate this threat, we performed a thorough cleaning step before using the data collected from LuCI.
In particular, we excluded failures that occur in a period where most of the test suite was failing for 15 consecutive builds in Linux and 9 builds in Windows. 
Additionally, we excluded failures that were deemed unreliable based on their history.
This allowed us to filter out 0.13\% of our initial dataset.

\section{Related Work}\label{section:related}

Flakiness is a well-known issue in Software Testing but research studies on this topic have only gained momentum in the past few years. Luo \etal~\cite{Luo2014} conducted the first empirical study on the root causes of flakiness. They analyzed commit fixes from 51 open-source projects and created the first taxonomy of flaky tests. Later on, several similar studies followed with different settings. Lam \etal~\cite{Lam2019RootCausing} conducted a study on flaky tests at Microsoft to identify and understand root causes of flaky tests.
They presented \textit{RootFinder}, a framework that helps to debug flaky tests using logs and time differences in their passing and failing runs. 
Romano \etal~\cite{romano2021empirical} focused their analysis on User Interface (UI) based tests. They showcased flakiness root causes for UI tests and highlighted the conditions needed to fix them. 
In the same vein, Gruber \etal \cite{Gruber2021} presented a large empirical analysis of more than 20,000 Python projects. They found test order-dependency and infrastructure to be among the top reasons for flakiness in those projects.

In addition to empirical analyses, tools have been introduced by researchers to help debugging, reproducing, and understanding flaky tests. 
Notably, \textit{DeFlaker}~\cite{Bell2018} detects flaky tests across commits, without performing any reruns, by checking for inconsistencies in test outcomes. More focused on test order-dependency, \textit{iDFlakies}~\cite{Lam2019iDFlakies} detects flaky tests by rerunning test suites in various orders.
To increase the chances of observing flakiness, Vysali \etal~\cite{ShakeIt} introduced \textit{Shaker}, a technique that relies on stress testing while rerunning potential flaky tests. 
As for fixing tools, Shi~\etal~\cite{Shi2019iFix} proposed iFixFlakies, a framework that recommends patches based on helpers present in other tests to automatically fix order-dependent flaky tests. 

While they remain scarce, the recent publication of datasets of flaky tests~\cite{Bell2018,Lam2019iDFlakies,Gruber2021} enabled new lines of work. Prediction models were suggested to differentiate flaky tests from non-flaky tests. King \etal~\cite{King2018} presented an approach that leverages Bayesian networks to classify and predict flaky tests based on code metrics. Pinto \etal~\cite{Pinto2020} used a bag of words representation of each test to train a model able to recognize flaky tests based on the vocabulary from the test code. This line of work has gained a lot of momentum lately as models achieved higher performances. Several works were carried out to replicate those studies and ensure their validity in different contexts~\cite{Haben2021,Camara2021VocabExtendedReplication}.
More recently, \textit{FlakeFlagger}~\cite{alshammari2021flakeflagger} has been introduced as another model using an extended set of features retrieved from the code under test and test smells. Compared to \textit{FlakeFlagger}, our approach uses a more lightweight set of features that are easily accessible after test executions, without the requirements of computation overhead --- \eg the requirement of code coverage.
Moreover, compared to the existing works which distinguish flaky tests from non-flaky tests, \fade{} focuses on test executions and discerns legitimate failures from flaky ones. 

Closer to our work, 
Herzig \etal~\cite{Herzig2015} used association rules to identify false alert patterns in the specific case of system and integration tests that contain steps. They evaluated their approach during the development of Windows 8.1 and Microsoft Dynamics AX.
They achieved a precision between 85\% and 90\% at detecting false testing alerts but had a low recall between 34\% and 48\%.
In our work, we cover a more varied set of tests, \ie GUI, integration, and unit tests.
Besides, we investigate a new set of dynamic properties (run properties and artefacts) that allow us to reach better performances (precision between 96\% and 88\% and a recall around 93\%).

In an industrial context, Kowalczyk \etal~\cite{Kowalczyk2020} implemented a flakiness scoring service at Apple. Their model quantifies the level of flakiness based on their historical flip rate and entropy (\ie changes in test outcomes across different revisions). Their goal was to identify and rank flaky tests to monitor and detect trends in flakiness. They were able to reduce flakiness by 44\% with less than 1\% loss in fault detection. 
In their assessment of test transitions, Leong \etal~\cite{LeongSPTM19} showed that flakiness can impact regression testing techniques. In particular, they showed that the strategy of transition-based prioritization, which reruns failing tests first, performs poorly because it automatically prioritizes flaky tests. These findings highlighted the need for testing solutions that take flakiness into consideration.

\section{Conclusion}\label{section:conclusion}
This paper addresses the rising issue of false testing alerts that mislead software developers and hinder the flow of continuous integration.
Our objective is to propose an approach for discerning legitimate failures from the lots of false alerts with high effectiveness and efficiency.
Therefore, we proposed \fade, a novel approach for failure classification using a lightweight set of features easily retrievable at runtime containing both dynamic properties (\eg run duration, run status, and logs) and static properties (\eg test source and command). 
To build and evaluate our approach, we relied on the continuous integration of the Chromium CI project.
We collected test failures over a period of 1\textonehalf  ~month and analyzed over 1 million test failures.
Our analysis shows that \fade can accurately detect legitimate failures even when considering different categories of tests: GUI, integration, and unit tests. 
The distinction between false alerts and legitimate failures capitalizes on failure properties like run duration and status but also uses the accompanying artefacts and spots keywords that are linked to flaky behaviour. 
Finally, we assessed the cost of \fade as an alternative to reruns for identifying legitimate failures. 
Once having a trained model --- which took about four minutes in our case --- \fade only requires 1 millisecond per build while reruns require 20 minutes.
To facilitate the reuse of our dataset and the replication of our study, we provide a comprehensive package including the used data and scripts\footnote{\url{https://github.com/GuillaumeHaben/FAIR-ReplicationPackage}}. 














\begin{acks}
This work is supported by the Facebook 2019 Testing and Verification research awards and PayPal.
\end{acks}

\bibliographystyle{ACM-Reference-Format}
\bibliography{0_main}


\begin{thebibliography}{30}


\ifx \showCODEN    \undefined \def \showCODEN     #1{\unskip}     \fi
\ifx \showDOI      \undefined \def \showDOI       #1{#1}\fi
\ifx \showISBNx    \undefined \def \showISBNx     #1{\unskip}     \fi
\ifx \showISBNxiii \undefined \def \showISBNxiii  #1{\unskip}     \fi
\ifx \showISSN     \undefined \def \showISSN      #1{\unskip}     \fi
\ifx \showLCCN     \undefined \def \showLCCN      #1{\unskip}     \fi
\ifx \shownote     \undefined \def \shownote      #1{#1}          \fi
\ifx \showarticletitle \undefined \def \showarticletitle #1{#1}   \fi
\ifx \showURL      \undefined \def \showURL       {\relax}        \fi
\providecommand\bibfield[2]{#2}
\providecommand\bibinfo[2]{#2}
\providecommand\natexlab[1]{#1}
\providecommand\showeprint[2][]{arXiv:#2}

\bibitem[\protect\citeauthoryear{Alshammari, Morris, Hilton, and
  Bell}{Alshammari et~al\mbox{.}}{2021}]%
        {alshammari2021flakeflagger}
\bibfield{author}{\bibinfo{person}{Abdulrahman Alshammari},
  \bibinfo{person}{Christopher Morris}, \bibinfo{person}{Michael Hilton}, {and}
  \bibinfo{person}{Jonathan Bell}.} \bibinfo{year}{2021}\natexlab{}.
\newblock \showarticletitle{Flakeflagger: Predicting flakiness without
  rerunning tests}. In \bibinfo{booktitle}{\emph{2021 IEEE/ACM 43rd
  International Conference on Software Engineering (ICSE)}}. IEEE,
  \bibinfo{pages}{1572--1584}.
\newblock


\bibitem[\protect\citeauthoryear{Bell, Legunsen, Hilton, Eloussi, Yung, and
  Marinov}{Bell et~al\mbox{.}}{2018}]%
        {Bell2018}
\bibfield{author}{\bibinfo{person}{Jonathan Bell}, \bibinfo{person}{Owolabi
  Legunsen}, \bibinfo{person}{Michael Hilton}, \bibinfo{person}{Lamyaa
  Eloussi}, \bibinfo{person}{Tifany Yung}, {and} \bibinfo{person}{Darko
  Marinov}.} \bibinfo{year}{2018}\natexlab{}.
\newblock \showarticletitle{{DeFlaker: Automatically Detecting Flaky Tests}}.
  In \bibinfo{booktitle}{\emph{Proceedings of the 40th International Conference
  on Software Engineering - ICSE '18}}. \bibinfo{publisher}{ACM Press},
  \bibinfo{address}{New York, New York, USA}, \bibinfo{pages}{433--444}.
\newblock
\showISBNx{9781450356381}
\urldef\tempurl%
\url{https://doi.org/10.1145/3180155.3180164}
\showDOI{\tempurl}


\bibitem[\protect\citeauthoryear{Bergstra and Bengio}{Bergstra and
  Bengio}{2012}]%
        {bergstra2012random}
\bibfield{author}{\bibinfo{person}{James Bergstra} {and}
  \bibinfo{person}{Yoshua Bengio}.} \bibinfo{year}{2012}\natexlab{}.
\newblock \showarticletitle{Random search for hyper-parameter optimization.}
\newblock \bibinfo{journal}{\emph{Journal of machine learning research}}
  \bibinfo{volume}{13}, \bibinfo{number}{2} (\bibinfo{year}{2012}).
\newblock


\bibitem[\protect\citeauthoryear{Camara, Silva, Endo, and Vergilio}{Camara
  et~al\mbox{.}}{2021}]%
        {Camara2021VocabExtendedReplication}
\bibfield{author}{\bibinfo{person}{B. Camara}, \bibinfo{person}{M. Silva},
  \bibinfo{person}{A.~T. Endo}, {and} \bibinfo{person}{S. Vergilio}.}
  \bibinfo{year}{2021}\natexlab{}.
\newblock \showarticletitle{What is the Vocabulary of Flaky Tests? An Extended
  Replication}. In \bibinfo{booktitle}{\emph{2021 2021 IEEE/ACM 29th
  International Conference on Program Comprehension (ICPC) (ICPC)}}.
  \bibinfo{publisher}{IEEE Computer Society}, \bibinfo{address}{Los Alamitos,
  CA, USA}, \bibinfo{pages}{444--454}.
\newblock
\urldef\tempurl%
\url{https://doi.org/10.1109/ICPC52881.2021.00052}
\showDOI{\tempurl}


\bibitem[\protect\citeauthoryear{Chicco and Jurman}{Chicco and Jurman}{2020}]%
        {chicco2020advantages}
\bibfield{author}{\bibinfo{person}{Davide Chicco} {and}
  \bibinfo{person}{Giuseppe Jurman}.} \bibinfo{year}{2020}\natexlab{}.
\newblock \showarticletitle{The advantages of the Matthews correlation
  coefficient (MCC) over F1 score and accuracy in binary classification
  evaluation}.
\newblock \bibinfo{journal}{\emph{BMC genomics}} \bibinfo{volume}{21},
  \bibinfo{number}{1} (\bibinfo{year}{2020}), \bibinfo{pages}{1--13}.
\newblock


\bibitem[\protect\citeauthoryear{contributors}{contributors}{2021}]%
        {TheChromiumProjects}
\bibfield{author}{\bibinfo{person}{Chromium contributors}.}
  \bibinfo{year}{2021}\natexlab{}.
\newblock \bibinfo{title}{The Chromium Projects}.
\newblock \bibinfo{howpublished}{\url{https://www.chromium.org/}}.
\newblock
\newblock
\shownote{(Accessed on 08/17/2021).}


\bibitem[\protect\citeauthoryear{Davis and Goadrich}{Davis and
  Goadrich}{2006}]%
        {precisionRecallCurve}
\bibfield{author}{\bibinfo{person}{Jesse Davis} {and} \bibinfo{person}{Mark
  Goadrich}.} \bibinfo{year}{2006}\natexlab{}.
\newblock \showarticletitle{The Relationship between Precision-Recall and ROC
  Curves}. In \bibinfo{booktitle}{\emph{Proceedings of the 23rd International
  Conference on Machine Learning}} (Pittsburgh, Pennsylvania, USA)
  \emph{(\bibinfo{series}{ICML '06})}. \bibinfo{publisher}{Association for
  Computing Machinery}, \bibinfo{address}{New York, NY, USA},
  \bibinfo{pages}{233–240}.
\newblock
\showISBNx{1595933832}
\urldef\tempurl%
\url{https://doi.org/10.1145/1143844.1143874}
\showDOI{\tempurl}


\bibitem[\protect\citeauthoryear{Dutta, Shi, Choudhary, Zhang, Jain, and
  Misailovic}{Dutta et~al\mbox{.}}{2020}]%
        {Dutta2020}
\bibfield{author}{\bibinfo{person}{Saikat Dutta}, \bibinfo{person}{August Shi},
  \bibinfo{person}{Rutvik Choudhary}, \bibinfo{person}{Zhekun Zhang},
  \bibinfo{person}{Aryaman Jain}, {and} \bibinfo{person}{Sasa Misailovic}.}
  \bibinfo{year}{2020}\natexlab{}.
\newblock \showarticletitle{{Detecting flaky tests in probabilistic and machine
  learning applications}}.
\newblock \bibinfo{journal}{\emph{ISSTA 2020 - Proceedings of the 29th ACM
  SIGSOFT International Symposium on Software Testing and Analysis}}
  (\bibinfo{year}{2020}), \bibinfo{pages}{211--224}.
\newblock
\showISBNx{9781450380089}
\urldef\tempurl%
\url{https://doi.org/10.1145/3395363.3397366}
\showDOI{\tempurl}


\bibitem[\protect\citeauthoryear{Eck, Castelluccio, Palomba, and Bacchelli}{Eck
  et~al\mbox{.}}{2019}]%
        {Eck2019}
\bibfield{author}{\bibinfo{person}{Moritz Eck}, \bibinfo{person}{Marco
  Castelluccio}, \bibinfo{person}{Fabio Palomba}, {and}
  \bibinfo{person}{Alberto Bacchelli}.} \bibinfo{year}{2019}\natexlab{}.
\newblock \showarticletitle{{Understanding Flaky Tests: The Developer's
  Perspective}}.
\newblock \bibinfo{journal}{\emph{arXiv}} (\bibinfo{year}{2019}),
  \bibinfo{pages}{830--840}.
\newblock
\showISBNx{9781450355728}


\bibitem[\protect\citeauthoryear{Goldberg}{Goldberg}{2017}]%
        {goldberg2017neural}
\bibfield{author}{\bibinfo{person}{Yoav Goldberg}.}
  \bibinfo{year}{2017}\natexlab{}.
\newblock \showarticletitle{Neural network methods for natural language
  processing}.
\newblock \bibinfo{journal}{\emph{Synthesis lectures on human language
  technologies}} \bibinfo{volume}{10}, \bibinfo{number}{1}
  (\bibinfo{year}{2017}), \bibinfo{pages}{1--309}.
\newblock


\bibitem[\protect\citeauthoryear{Gruber, Lukasczyk, Krois, and Fraser}{Gruber
  et~al\mbox{.}}{2021}]%
        {Gruber2021}
\bibfield{author}{\bibinfo{person}{Martin Gruber}, \bibinfo{person}{Stephan
  Lukasczyk}, \bibinfo{person}{Florian Krois}, {and} \bibinfo{person}{Gordon
  Fraser}.} \bibinfo{year}{2021}\natexlab{}.
\newblock \showarticletitle{{An Empirical Study of Flaky Tests in Python}}.
\newblock \bibinfo{journal}{\emph{Proceedings - 2021 IEEE 14th International
  Conference on Software Testing, Verification and Validation, ICST 2021}}
  (\bibinfo{year}{2021}), \bibinfo{pages}{148--158}.
\newblock
\showISBNx{9781728168364}
\urldef\tempurl%
\url{https://doi.org/10.1109/ICST49551.2021.00026}
\showDOI{\tempurl}


\bibitem[\protect\citeauthoryear{Haben, Habchi, Papadakis, Cordy, and {Le
  Traon}}{Haben et~al\mbox{.}}{2021}]%
        {Haben2021}
\bibfield{author}{\bibinfo{person}{Guillaume Haben}, \bibinfo{person}{Sarra
  Habchi}, \bibinfo{person}{Mike Papadakis}, \bibinfo{person}{Maxime Cordy},
  {and} \bibinfo{person}{Yves {Le Traon}}.} \bibinfo{year}{2021}\natexlab{}.
\newblock \showarticletitle{{A Replication Study on the Usability of Code
  Vocabulary in Predicting Flaky Tests}}.
\newblock \bibinfo{journal}{\emph{Proceedings of the International Conference
  on Mining Software Repositories (MSR)}} (\bibinfo{year}{2021}).
\newblock


\bibitem[\protect\citeauthoryear{Herzig and Nagappan}{Herzig and
  Nagappan}{2015}]%
        {Herzig2015}
\bibfield{author}{\bibinfo{person}{Kim Herzig} {and}
  \bibinfo{person}{Nachiappan Nagappan}.} \bibinfo{year}{2015}\natexlab{}.
\newblock \showarticletitle{{Empirically Detecting False Test Alarms Using
  Association Rules}}.
\newblock \bibinfo{journal}{\emph{Proceedings - International Conference on
  Software Engineering}}  \bibinfo{volume}{2} (\bibinfo{year}{2015}),
  \bibinfo{pages}{39--48}.
\newblock
\showISBNx{9781479919345}
\showISSN{02705257}
\urldef\tempurl%
\url{https://doi.org/10.1109/ICSE.2015.133}
\showDOI{\tempurl}


\bibitem[\protect\citeauthoryear{King, Santiago, Phillips, and Clarke}{King
  et~al\mbox{.}}{2018}]%
        {King2018}
\bibfield{author}{\bibinfo{person}{Tariq~M King}, \bibinfo{person}{Dionny
  Santiago}, \bibinfo{person}{Justin Phillips}, {and} \bibinfo{person}{Peter~J
  Clarke}.} \bibinfo{year}{2018}\natexlab{}.
\newblock \showarticletitle{{Towards a Bayesian Network Model for Predicting
  Flaky Automated Tests}}.
\newblock \bibinfo{journal}{\emph{2018 IEEE International Conference on
  Software Quality, Reliability and Security Companion (QRS-C)}}
  (\bibinfo{year}{2018}), \bibinfo{pages}{100--107}.
\newblock
\showISBNx{9781538678398}
\urldef\tempurl%
\url{https://doi.org/10.1109/QRS-C.2018.00031}
\showDOI{\tempurl}


\bibitem[\protect\citeauthoryear{Kowalczyk, Nair, Gao, Silberstein, Long, and
  Memon}{Kowalczyk et~al\mbox{.}}{2020}]%
        {Kowalczyk2020}
\bibfield{author}{\bibinfo{person}{Emily Kowalczyk}, \bibinfo{person}{Karan
  Nair}, \bibinfo{person}{Zebao Gao}, \bibinfo{person}{Leo Silberstein},
  \bibinfo{person}{Teng Long}, {and} \bibinfo{person}{Atif Memon}.}
  \bibinfo{year}{2020}\natexlab{}.
\newblock \showarticletitle{Modeling and Ranking Flaky Tests at Apple}. In
  \bibinfo{booktitle}{\emph{Proceedings of the ACM/IEEE 42nd International
  Conference on Software Engineering: Software Engineering in Practice}}
  (Seoul, South Korea) \emph{(\bibinfo{series}{ICSE-SEIP '20})}.
  \bibinfo{publisher}{Association for Computing Machinery},
  \bibinfo{address}{New York, NY, USA}, \bibinfo{pages}{110–119}.
\newblock
\showISBNx{9781450371230}
\urldef\tempurl%
\url{https://doi.org/10.1145/3377813.3381370}
\showDOI{\tempurl}


\bibitem[\protect\citeauthoryear{Lam, Godefroid, Nath, Santhiar, and
  Thummalapenta}{Lam et~al\mbox{.}}{2019a}]%
        {Lam2019RootCausing}
\bibfield{author}{\bibinfo{person}{Wing Lam}, \bibinfo{person}{Patrice
  Godefroid}, \bibinfo{person}{Suman Nath}, \bibinfo{person}{Anirudh Santhiar},
  {and} \bibinfo{person}{Suresh Thummalapenta}.}
  \bibinfo{year}{2019}\natexlab{a}.
\newblock \showarticletitle{{Root Causing Flaky Tests in a Large-Scale
  Industrial Setting}}. In \bibinfo{booktitle}{\emph{Proceedings ofthe 28th ACM
  SIGSOFT International Symposium on Software Testing and Analysis (ISSTA
  '19)}}. \bibinfo{publisher}{ACM Press}, \bibinfo{address}{Beijing, China},
  \bibinfo{pages}{101--111}.
\newblock
\showISBNx{9781450362245}
\urldef\tempurl%
\url{https://doi.org/10.1145/3293882.3330570}
\showDOI{\tempurl}


\bibitem[\protect\citeauthoryear{Lam, Oei, Shi, Marinov, and Xie}{Lam
  et~al\mbox{.}}{2019b}]%
        {Lam2019iDFlakies}
\bibfield{author}{\bibinfo{person}{Wing Lam}, \bibinfo{person}{Reed Oei},
  \bibinfo{person}{August Shi}, \bibinfo{person}{Darko Marinov}, {and}
  \bibinfo{person}{Tao Xie}.} \bibinfo{year}{2019}\natexlab{b}.
\newblock \showarticletitle{{IDFlakies: A framework for detecting and partially
  classifying flaky tests}}.
\newblock \bibinfo{journal}{\emph{Proceedings - 2019 IEEE 12th International
  Conference on Software Testing, Verification and Validation, ICST 2019}}
  (\bibinfo{year}{2019}), \bibinfo{pages}{312--322}.
\newblock
\showISBNx{9781728117355}
\urldef\tempurl%
\url{https://doi.org/10.1109/ICST.2019.00038}
\showDOI{\tempurl}


\bibitem[\protect\citeauthoryear{Lam, Winter, Wei, Xie, Marinov, and Bell}{Lam
  et~al\mbox{.}}{2020}]%
        {Lam2020b}
\bibfield{author}{\bibinfo{person}{Wing Lam}, \bibinfo{person}{Stefan Winter},
  \bibinfo{person}{Anjiang Wei}, \bibinfo{person}{Tao Xie},
  \bibinfo{person}{Darko Marinov}, {and} \bibinfo{person}{Jonathan Bell}.}
  \bibinfo{year}{2020}\natexlab{}.
\newblock \showarticletitle{{A large-scale longitudinal study of flaky tests}}.
\newblock \bibinfo{journal}{\emph{Proceedings of the ACM on Programming
  Languages}} \bibinfo{volume}{4}, \bibinfo{number}{OOPSLA}
  (\bibinfo{year}{2020}), \bibinfo{pages}{1--29}.
\newblock
\showISSN{2475-1421}
\urldef\tempurl%
\url{https://doi.org/10.1145/3428270}
\showDOI{\tempurl}


\bibitem[\protect\citeauthoryear{Leong, Singh, Papadakis, Traon, and
  Micco}{Leong et~al\mbox{.}}{2019}]%
        {LeongSPTM19}
\bibfield{author}{\bibinfo{person}{Claire Leong}, \bibinfo{person}{Abhayendra
  Singh}, \bibinfo{person}{Mike Papadakis}, \bibinfo{person}{Yves~Le Traon},
  {and} \bibinfo{person}{John Micco}.} \bibinfo{year}{2019}\natexlab{}.
\newblock \showarticletitle{Assessing transition-based test selection
  algorithms at Google}. In \bibinfo{booktitle}{\emph{Proceedings of the 41st
  International Conference on Software Engineering: Software Engineering in
  Practice, {ICSE} {(SEIP)} 2019, Montreal, QC, Canada, May 25-31, 2019}},
  \bibfield{editor}{\bibinfo{person}{Helen Sharp} {and} \bibinfo{person}{Mike
  Whalen}} (Eds.). \bibinfo{publisher}{{IEEE} / {ACM}},
  \bibinfo{pages}{101--110}.
\newblock
\urldef\tempurl%
\url{https://doi.org/10.1109/ICSE-SEIP.2019.00019}
\showDOI{\tempurl}


\bibitem[\protect\citeauthoryear{Luo, Hariri, Eloussi, and Marinov}{Luo
  et~al\mbox{.}}{2014}]%
        {Luo2014}
\bibfield{author}{\bibinfo{person}{Qingzhou Luo}, \bibinfo{person}{Farah
  Hariri}, \bibinfo{person}{Lamyaa Eloussi}, {and} \bibinfo{person}{Darko
  Marinov}.} \bibinfo{year}{2014}\natexlab{}.
\newblock \showarticletitle{{An empirical analysis of flaky tests}}. In
  \bibinfo{booktitle}{\emph{Proceedings of the ACM SIGSOFT Symposium on the
  Foundations of Software Engineering}},
  Vol.~\bibinfo{volume}{16-21-November-2014}. \bibinfo{publisher}{Association
  for Computing Machinery}, \bibinfo{pages}{643--653}.
\newblock
\showISBNx{9781450330565}
\urldef\tempurl%
\url{https://doi.org/10.1145/2635868.2635920}
\showDOI{\tempurl}


\bibitem[\protect\citeauthoryear{Micco}{Micco}{2017}]%
        {Micco2017}
\bibfield{author}{\bibinfo{person}{John Micco}.}
  \bibinfo{year}{2017}\natexlab{}.
\newblock \bibinfo{title}{{The State of Continuous Integration Testing
  Google}}.
\newblock
\newblock


\bibitem[\protect\citeauthoryear{Palmer}{Palmer}{2019}]%
        {FlakinessSpotify}
\bibfield{author}{\bibinfo{person}{Jason Palmer}.}
  \bibinfo{year}{2019}\natexlab{}.
\newblock \bibinfo{title}{Test Flakiness – Methods for identifying and
  dealing with flaky tests : Spotify Engineering}.
\newblock
  \bibinfo{howpublished}{\url{https://engineering.atspotify.com/2019/11/18/test-flakiness-methods-for-identifying-and-dealing-with-flaky-tests/}}.
\newblock
\newblock
\shownote{(Accessed on 01/12/2021).}


\bibitem[\protect\citeauthoryear{Pinto, Miranda, Dissanayake, D'Amorim, Treude,
  and Bertolino}{Pinto et~al\mbox{.}}{2020}]%
        {Pinto2020}
\bibfield{author}{\bibinfo{person}{Gustavo Pinto}, \bibinfo{person}{Breno
  Miranda}, \bibinfo{person}{Supun Dissanayake}, \bibinfo{person}{Marcelo
  D'Amorim}, \bibinfo{person}{Christoph Treude}, {and} \bibinfo{person}{Antonia
  Bertolino}.} \bibinfo{year}{2020}\natexlab{}.
\newblock \showarticletitle{{What is the Vocabulary of Flaky Tests?}}
\newblock \bibinfo{journal}{\emph{Proceedings - 2020 IEEE/ACM 17th
  International Conference on Mining Software Repositories, MSR 2020}}
  (\bibinfo{year}{2020}), \bibinfo{pages}{492--502}.
\newblock
\showISBNx{9781450379571}
\urldef\tempurl%
\url{https://doi.org/10.1145/3379597.3387482}
\showDOI{\tempurl}


\bibitem[\protect\citeauthoryear{Rahman and Rigby}{Rahman and Rigby}{2018}]%
        {Rahman2018}
\bibfield{author}{\bibinfo{person}{M.~Tajmilur Rahman} {and}
  \bibinfo{person}{Peter~C. Rigby}.} \bibinfo{year}{2018}\natexlab{}.
\newblock \showarticletitle{{The impact of failing, flaky, and high failure
  tests on the number of crash reports associated with firefox builds}}.
\newblock \bibinfo{journal}{\emph{ESEC/FSE 2018 - Proceedings of the 2018 26th
  ACM Joint Meeting on European Software Engineering Conference and Symposium
  on the Foundations of Software Engineering}} (\bibinfo{year}{2018}),
  \bibinfo{pages}{857--862}.
\newblock
\showISBNx{9781450355735}
\urldef\tempurl%
\url{https://doi.org/10.1145/3236024.3275529}
\showDOI{\tempurl}


\bibitem[\protect\citeauthoryear{Rehkopf}{Rehkopf}{[n.d.]}]%
        {CI}
\bibfield{author}{\bibinfo{person}{Max Rehkopf}.}
  \bibinfo{year}{[n.d.]}\natexlab{}.
\newblock \bibinfo{title}{What is Continuous Integration | Atlassian}.
\newblock
  \bibinfo{howpublished}{\url{https://www.atlassian.com/continuous-delivery/continuous-integration}}.
\newblock
\newblock
\shownote{(Accessed on 01/12/2021).}


\bibitem[\protect\citeauthoryear{Romano, Song, Grandhi, Yang, and Wang}{Romano
  et~al\mbox{.}}{2021}]%
        {romano2021empirical}
\bibfield{author}{\bibinfo{person}{Alan Romano}, \bibinfo{person}{Zihe Song},
  \bibinfo{person}{Sampath Grandhi}, \bibinfo{person}{Wei Yang}, {and}
  \bibinfo{person}{Weihang Wang}.} \bibinfo{year}{2021}\natexlab{}.
\newblock \showarticletitle{An Empirical Analysis of UI-based Flaky Tests}. In
  \bibinfo{booktitle}{\emph{2021 IEEE/ACM 43rd International Conference on
  Software Engineering (ICSE)}}. IEEE, \bibinfo{pages}{1585--1597}.
\newblock


\bibitem[\protect\citeauthoryear{Shi, Lam, Oei, Xie, and Marinov}{Shi
  et~al\mbox{.}}{2019}]%
        {Shi2019iFix}
\bibfield{author}{\bibinfo{person}{August Shi}, \bibinfo{person}{Wing Lam},
  \bibinfo{person}{Reed Oei}, \bibinfo{person}{Tao Xie}, {and}
  \bibinfo{person}{Darko Marinov}.} \bibinfo{year}{2019}\natexlab{}.
\newblock \showarticletitle{{iFixFlakies : A Framework for Automatically Fixing
  Order-Dependent Flaky Tests}}. In \bibinfo{booktitle}{\emph{27th ACM Joint
  European Software Engineering Conference and Symposium on the Foundations
  ofSoftware Engineering (ESEC/FSE '19)}}.
\newblock
\showISBNx{9781450355728}
\urldef\tempurl%
\url{https://doi.org/10.1145/3338906.3338925}
\showDOI{\tempurl}


\bibitem[\protect\citeauthoryear{team}{team}{2021}]%
        {onlineChromiumGithub}
\bibfield{author}{\bibinfo{person}{The Chromium~Development team}.}
  \bibinfo{year}{2021}\natexlab{}.
\newblock \bibinfo{title}{chromium/chromium: The official GitHub mirror of the
  Chromium source}.
\newblock \bibinfo{howpublished}{\url{https://github.com/chromium/chromium}}.
\newblock
\newblock
\shownote{(Accessed on 07/06/2021).}


\bibitem[\protect\citeauthoryear{Thorve, Sreshtha, and Meng}{Thorve
  et~al\mbox{.}}{2018}]%
        {Thorve2018}
\bibfield{author}{\bibinfo{person}{Swapna Thorve}, \bibinfo{person}{Chandani
  Sreshtha}, {and} \bibinfo{person}{Na Meng}.} \bibinfo{year}{2018}\natexlab{}.
\newblock \showarticletitle{{An empirical study of flaky tests in android
  apps}}.
\newblock \bibinfo{journal}{\emph{Proceedings - 2018 IEEE International
  Conference on Software Maintenance and Evolution, ICSME 2018}}
  (\bibinfo{year}{2018}), \bibinfo{pages}{534--538}.
\newblock
\showISBNx{9781538678701}
\urldef\tempurl%
\url{https://doi.org/10.1109/ICSME.2018.00062}
\showDOI{\tempurl}


\bibitem[\protect\citeauthoryear{Vaidhyam~Subramanian, McIntosh, and
  Adams}{Vaidhyam~Subramanian et~al\mbox{.}}{2020}]%
        {ShakeIt}
\bibfield{author}{\bibinfo{person}{Shivashree~Vysali Vaidhyam~Subramanian},
  \bibinfo{person}{Shane McIntosh}, {and} \bibinfo{person}{Bram Adams}.}
  \bibinfo{year}{2020}\natexlab{}.
\newblock \showarticletitle{Quantifying, Characterizing, and Mitigating Flakily
  Covered Program Elements}.
\newblock \bibinfo{journal}{\emph{IEEE Transactions on Software Engineering}}
  (\bibinfo{year}{2020}), \bibinfo{pages}{1--1}.
\newblock
\urldef\tempurl%
\url{https://doi.org/10.1109/TSE.2020.3010045}
\showDOI{\tempurl}


\end{thebibliography}

\end{document}